\documentclass{PoS}
\usepackage{amsmath}
\title{$\Delta I=3/2$, $K\to\pi\pi$ Decays with Light, Non-Zero Momentum Pions}

\ShortTitle{$\Delta I=3/2$, $K\to\pi\pi$ Decays with Light, Non-Zero Momentum Pions}

\author{\speaker{Matthew Lightman}\\
        Department of Physics, Columbia University, New York, NY 10027, USA\\
        E-mail: \email{lightman@phys.columbia.edu}}

\author{Elaine Goode\\
        University of Southampton, School of Physics and Astronomy, Highfield, Southampton, SO17 1BJ, United Kingdom\\
	E-mail: \email{ejg4g08@soton.ac.uk}}

\author{RBC and UKQCD collaborations}

\abstract{$\Delta I=3/2$, $K\to\pi\pi$ matrix elements are calculated on 68 configurations of quenched $24^3\times 64$ lattices using the DBW2 action, and domain wall fermions with $L_s=16$.  The lattice spacing is $a^{-1}=1.3\text{ GeV}$, corresponding to a physical volume of $(3.6\text{ fm})^3$, which allows us to simulate a pion mass of $m_\pi=227.6(6)\text{ MeV}$ and a kaon mass of $m_K=564(2)\text{ MeV}$.  Twisted boundary conditions are used to give the two pions momentum.  One twist corresponds to a pion momentum of $p_\pi=\pi/L=170\text{ MeV}$, which represents a decay that is nearly on-shell.  Results for time separations of 20, 24, 28, and 32 between the kaon and the two pions are computed and an error weighted average is performed to reduced the error.  The matrix elements are then found to have errors of order 3-4\% for momentum 0 and $\pi/L$, 7\% for momentum $\sqrt{2}\pi/L$, and 15\% for momentum $\sqrt{3}\pi/L$.}

\FullConference{The XXVII International Symposium on Lattice Field Theory - LAT2009\\
		 July 26-31 2009\\
		 Peking University, Beijing, China}

\begin{document}

\vspace{-0.15 in}
\section{Introduction}
\vspace{-0.1 in}

Interest in precise lattice calculations of $K\to\pi\pi$ decays stems from the possibility of gaining insight into the origin of the $\Delta I=1/2$ rule, and of putting constraints on CKM matrix elements.  In particular, $\epsilon'/\epsilon$ and the CP violating phase can be computed when lattice calculations are compared with experimental results \cite{CPPACS,RBC}.

Several quenched $K\to\pi\pi$ calculations have been performed in the past \cite{RBC,Changhoan_thesis,Changhoan1,Changhoan2,Takeshi}.  The calculation presented here is a quenched calculation using domain wall fermions (DWF) on a $24^3\times 64$, $L_s=16$ lattice with strong coupling ($a^{-1}=1.3\text{ GeV}$).  This represents a larger spatial volume, $(3.6\text{ fm})^3$, than in previous works, allowing for a lighter, near physical pion mass of $m_\pi=227.6(6)\text{ MeV}$.  As in Refs. \cite{Changhoan_thesis,Changhoan1,Changhoan2,Takeshi} we are directly calculating the two pion decay amplitudes and not relying on chiral perturbation theory to determine these amplitudes from the simpler $K\to\pi$ and $K\to |0\rangle$ amplitudes.  This use of chiral perturbation theory, upon which Refs. \cite{CPPACS,RBC} were based, now appears to introduce large, uncontrolled systematic errors \cite{Christ_Lat_08}.  This quenched study is a pilot project that will be followed by a full QCD calculation using the RBC/UKQCD $32^3\times 64$, $L_s=32$ lattices with 2+1 flavors of DWF and strong coupling ($a^{-1}=1.4$ GeV) that are currently being generated, with light pion masses ($m_\pi \le 250$ MeV).

\vspace{-0.15 in}
\section{Four Quark Operators and the Effective Hamiltonian}
\vspace{-0.1 in}

The weak interactions and the effects of the heavier quarks can be included in the lattice QCD simulation by evaluating matrix elements of an effective Hamiltonian \cite{Ciuchini,Buchalla}.  In particular we use the conventions of equation 3 in \cite{RBC}.  In this formalism we are interested in calculating matrix elements of four quark operators between $K$ and $\pi\pi$ states.  Here we calculate matrix elements with $\Delta I=3/2$ operators, where $\Delta I$ is the change in isospin.  The $\Delta I=3/2$ operators are further classified by how they transform under $SU(3)_L\times SU(3)_R$ into ``(27,1)'', ``(8,8)'' and ``(8,8) mixed'' parts. \cite{CPPACS,RBC}.

\vspace{-0.15 in}
\section{Twisted Boundary Conditions}
\vspace{-0.1 in}

The kinematics of the physical $K\to\pi\pi$ decay are such that the pions have significant momentum, $\sim 200$ MeV each, so that it is required to simulate a two pion excited state on the lattice.  Extracting an excited state requires multi-exponential fitting making it difficult to obtain a signal.  One remedy is to use \emph{twisted boundary conditions} \cite{Changhoan_thesis,Changhoan1,Changhoan2,Sachrajda_Villadoro}, in which the boundary conditions on a quark field, instead of simply being periodic, cause it to change by a phase $e^{i\phi}$ when going through the boundary.  We say that this spatial direction is ``twisted'' by an amount $\phi$.

Twisting a quark by $\phi$ will cause it to have momentum
\begin{equation}
p_n=\frac{\phi+2\pi n}{L}
\end{equation}
where $L$ is the spatial extent of the lattice.  In particular, we use twists of amount $\pi$ (corresponding to \emph{antiperiodic} boundary conditions in the twisted direction) to obtain a two pion system with 0 total momentum.  For example, if only the $x$ direction is twisted then we can have a two pion ground state in which one pion has momentum $p_x=\frac{\pi}{L}$ and the other pion has momentum $p_x=-\frac{\pi}{L}$.  We can obtain two pion ground states with pion 3-momentum of magnitude
\begin{equation}
p_\pi=\frac{\pi}{L},\frac{\sqrt{2}\pi}{L},\frac{\sqrt{3}\pi}{L}
\end{equation}
by twisting one, two, and three spatial directions respectively.  Since we actually perform the calculation for the decay $K^+\to\pi^+\pi^+$, and relate it to the physical decay via the Wigner-Eckhart theorem \cite{Changhoan_thesis}, we can twist the down quark only, thus giving the two pions momentum and giving no momentum to the kaon.  By rotational symmetry we can average over results for pion momenta of the same magnitude but different directions, i.e. we can average over results with the same number of twists.

\vspace{-0.15 in}
\section{Details of the Calculation}
\vspace{-0.1 in}

The calculation was carried out on 68 configurations of quenched $24^3\times 64$ lattices using the DBW2 action, and domain wall fermions with $L_s=16$.  We used the QCDOC computers at Brookhaven and Columbia University for both configuration generation and propagator inversion.  The inverse lattice spacing is $a^{-1}=1.3\text{ GeV}$, the physical volume is $(3.6\text{ fm})^3$, and we set the light and strange quark masses to $m_l=0.0055$ (chosen so that $m_\pi L\approx 4$) and $m_s=0.08$ in lattice units respectively.  This yielded a pion mass of $m_\pi=227.6(6)\text{ MeV}$ and a kaon mass of $m_K=564(2)\text{ MeV}$ in physical units.  The pion momentum corresponding to one twist for these lattices is $p_\pi=\pi/L=170\text{ MeV}$ so that the decay is expected to be nearly energy conserving ($E_{\pi\pi}-m_K\approx 2\sqrt{m_\pi^2+p_\pi^2}-m_K=4\text{ MeV}$).

We added and subtracted quark propagators with periodic and antiperiodic boundary conditions in the \emph{time} direction from each other in order to double the effective time length.  Periodic plus antiperiodic (P+A) propagators thus provide a source at $t=0$ and periodic minus antiperiodic (P-A) propagators provide an effective source at $t=64$.  These provide the left and right walls for the kaon and two pions and the time $t$ of the four quark operator is varied.  The kaon mass, two pion energy, and $K\to\pi\pi$ matrix element were extracted from the asymptotic behavior of the corresponding correlation functions as in \cite{My_Lat_08}, the only difference being that the $K\to\pi\pi$ correlator was fit directly rather than fitting the quotient of correlators defined in equation 3.5 of \cite{My_Lat_08}.

For the kaon and the two pions with zero momentum we used propagators with Coulomb gauge fixed wall sources.  For the two pions with non-zero momentum, we used the same type of propagators for the $u$ quark, but used propagators with twisted (antiperiodic spatial) boundary conditions for the $d$ quark with Coulomb gauge fixed momentum wall sources of the ``cosine'' type
\begin{equation}\label{cosine_src}
s_{{\bf p},\text{cosine}}({\bf x})=\cos(p_x x)\cos(p_y y)\cos(p_z z)
\end{equation}

Had we used sources of the ``pure momentum'' type
\begin{equation}
s_{\bf p}({\bf x})=e^{i{\bf p}\cdot {\bf x}} .
\end{equation}
it would have been necessary to do an inversion with source momentum $+{\bf p}$ for one $d$ quark and an inversion with source momentum $-{\bf p}$ for the other $d$ quark in order to couple to a two pion state with zero total momentum.  Using the same cosine source propagator for each $d$ quark eliminates the need for this extra inversion, but this leads to cross terms that couple to two pion states with non-zero total momentum.  For example, with a single twist in the $x$ direction, we hope to couple to a two pion state with individual pion momenta ${\bf p}_1=-\frac{\pi}{L}{\bf \hat{x}}$ and ${\bf p}_2=\frac{\pi}{L}{\bf \hat{x}}$, and thus the product of the sources of the two $d$ quarks according to Equation \ref{cosine_src} should be
\begin{align}
s_{{\bf p}_1,\text{cosine}}({\bf x}_1)s_{{\bf p}_2,\text{cosine}}({\bf x}_2)&=\cos\left(\frac{\pi}{L}x_1\right)\cos\left(\frac{\pi}{L}x_2\right) \nonumber \\
&=\frac{1}{4}\left(e^{i\frac{\pi}{L}x_1}e^{i\frac{\pi}{L}x_2}+e^{i\frac{\pi}{L}x_1}e^{-i\frac{\pi}{L}x_2}+e^{-i\frac{\pi}{L}x_1}e^{i\frac{\pi}{L}x_2}+e^{-i\frac{\pi}{L}x_1}e^{-i\frac{\pi}{L}x_2}\right) \label{src_prod}
\end{align}
But in Equation \ref{src_prod} the first and last term couple to two pion states with total momentum $2\frac{\pi}{L}$ and $-2\frac{\pi}{L}$ in the $x$ direction respectively.  To circumvent this difficulty, we calculate two pion correlators with cosine sources but pure momentum sinks.  The unwanted terms from the source vanish in the limit of an infinite number of configurations since the two pion final state is constrained to have zero total momentum by the pure momentum sink, but since these are sinks they require no extra inversions.  In the $K\to\pi\pi$ correlator, the 0 momentum kaon in the initial state has a similar effect on the cosine sources of the two pions in the final state.

\vspace{-0.15 in}
\section{Results of the First Calculation}
\vspace{-0.1 in}

The calculation was carried out as described, and the results for the matrix elements are shown in Table \ref{tb:bad_data}.  The errors on the (8,8) and (8,8) \emph{mixed} matrix elements with momentum $\pi/L$ (where $L=24$) are of order 50\%, and the data was too noisy to extract a signal for the (27,1) matrix element with momentum $\pi/L$ and for all the matrix elements with momentum $\sqrt{2}\pi/L$ and $\sqrt{3}\pi/L$.  This is due to the fact that a time separation of 64 is too large, especially for the higher momenta where the higher energy of the two pion state causes its signal to decay into noise well before it can reach the four quark operator.

\begin{table}
\caption{Lattice matrix elements for a time separation of 64 between the kaons and the two pions (first calculation).  Note that these values differ from those presented in the talk where a normalization factor was omitted.}\label{tb:bad_data}
\begin{center}
\begin{tabular}{|c|c|c|c|c|c|c|}
\hline
$p$&$\mathcal{M}_{(27,1)}$&$\mathcal{M}_{(8,8)}$&$\mathcal{M}_{(8,8)\text{ mixed}}$\\
\hline
0&0.00408(40)&0.0950(77)&0.294(34)\\
$\pi/L$&-&0.161(85)&0.55(30)\\
\hline
\end{tabular}
\end{center}
\end{table}

\vspace{-0.15 in}
\section{Results of the Second Calculation}
\vspace{-0.1 in}

The calculation was redone with smaller time separations between the kaon and the two pions.\footnote{These results were not presented in the talk.}  Specifically, propagators with sources at $t_K=$20, 24, 28, 32, 36, 40, and 44 were generated in order to calculate $K\to\pi\pi$ correlators with kaon sources at these times, while the two pion sources remained at either the left or the right walls.  Thus we could achieve time separations between the kaon and the two pions of 20, 24, 28, and 32 in two different ways in order to double the statistics (for example a time separation of 20 results when the two pions are at the left wall, i.e. $t_\pi=0$, and when $t_K=20$, and also results when the two pions are at the right wall, i.e. $t_\pi=64$, and when $t_K=44$).  These separations were chosen so that the kaon and two pion sources were close enough to the four quark operator (with the operator between the sources) so that the signals from the sources do not decay into noise before reaching the operator, but were sufficiently distant from each other that a plateau region can be found where the effects of possible excited states have died away.

\begin{figure}[htp]
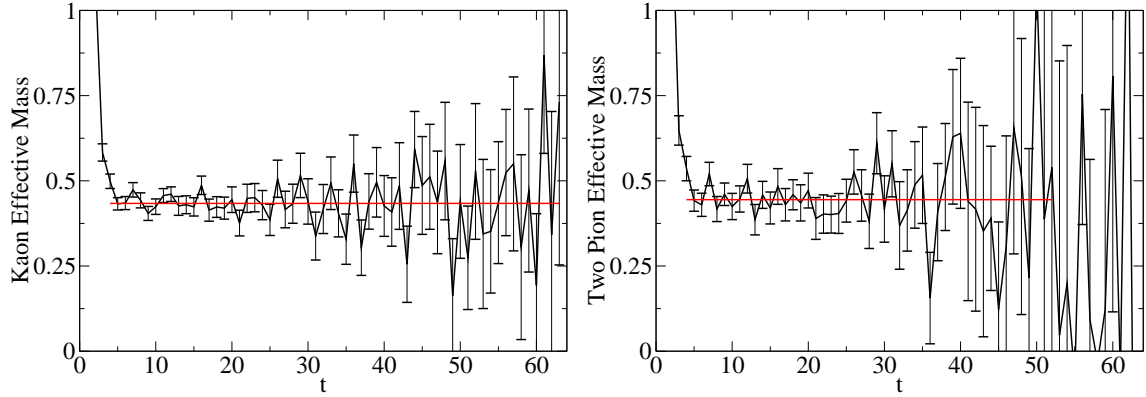

\centering
\includegraphics[scale=0.3,clip=true]{kaon_meff_comb}
\hspace{0.02 in}
\includegraphics[scale=0.3,clip=true]{pipi_meff_comb_momnum1}
\hfill
\caption{Effective mass plots for the kaon correlator (left) and the two pion correlator for pions with momentum $\pi/L$ (right).  The effective mass of a correlator C(t) is defined as $m_\mathrm{eff}(t)=-\ln\left(C(t)/C(t-1)\right)$.}\label{fig:kaon_pipi_meff}
\end{figure}

\begin{figure}[htp]
\centering
\includegraphics[scale=0.4,clip=true]{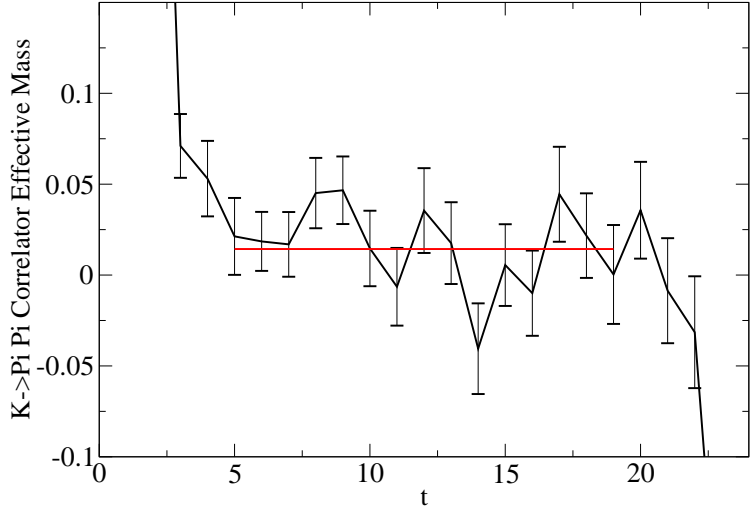}
\hfill
\caption{Effective mass plot for the $K\to\pi\pi$ correlator with the (27,1) four quark operator, pions with momentum $\pi/L$, the two pion source located at $t=0$ and the kaon source located at $t=24$.}\label{fig:kpipi_meff}
\end{figure}

Effective mass plots for the kaon correlator, the two pion correlator with momentum $\pi/L$, and the $K\to\pi\pi$ correlator with the (27,1) operator, momentum $\pi/L$ and time separation 24 between the two sources, are shown in Figures \ref{fig:kaon_pipi_meff} and \ref{fig:kpipi_meff}.  The two pion energies that were extracted from these effective mass plots are given in Table \ref{tb:EPiPi}.  Table \ref{tb:Mvals} shows the lattice matrix elements obtained for different time separations of the two sources, as well as an error weighted average of the results from the four different time separations.  The error of the error weighted average is smaller than the error of the matrix element for any single time separation in all cases.  For momentum 0 and $\pi/L$ the error of the error weighted average is of order 3-4\% (for all of the operators), for momentum $\sqrt{2}\pi/L$ it is of order 7\%, and for momentum $\sqrt{3}\pi/L$ it is of order 15\%.

\begin{table}
\caption{Values of the two pion energies ($E_{\pi\pi}$) extracted from the effective mass plots.  These are compared to the value one would expect in the absence of interactions between the two pions, $2\sqrt{m_\pi^2+p^2}$, as a sanity check.}\label{tb:EPiPi}
\begin{center}
\begin{tabular}{|c|c|c|c|c|}
\hline
&$2\sqrt{m_\pi^2+p^2}$ (lattice units)&$E_{\pi\pi}$ (lattice units)&$2\sqrt{m_\pi^2+p^2}$ (MeV)&$E_{\pi\pi}$ (MeV)\\
\hline
$p=0$&0.3501&0.3509(13)&455&456(2)\\
$p=\pi/L$&0.4372&0.4447(19)&568&578(2)\\
$p=\sqrt{2}\pi/L$&0.5096&0.5316(50)&662&691(6)\\
$p=\sqrt{3}\pi/L$&0.5729&0.577(20)&745&750(27)\\
\hline
\end{tabular}
\end{center}
\end{table}

\begin{table}
\caption{Lattice matrix elements for the different time separations between the kaon and the two pions, $\Delta t=$20, 24, 28, and 32.  The error weighted average of the results from these four separations is shown in the final column.  Notice that its error is generally significantly smaller than the corresponding errors in the other columns.}\label{tb:Mvals}
\begin{center}
\begin{tabular}{|c|c|c|c|c|c|}
\hline
\multicolumn{6}{|c|}{Matrix Elements with (27,1) Operator} \\
\hline
&$\Delta t=20$&$\Delta t=24$&$\Delta t=28$&$\Delta t=32$&Error Weighted Ave.\\
\hline
$p=0$&0.00529(29)&0.00540(28)&0.00490(26)&0.00550(34)&0.00526(17)\\
$p=\pi/L$&0.00882(55)&0.00923(59)&0.00813(57)&0.00924(68)&0.00884(37)\\
$p=\sqrt{2}\pi/L$&0.0140(11)&0.0150(16)&0.0137(17)&0.0157(19)&0.01449(99)\\
$p=\sqrt{3}\pi/L$&0.0249(47)&0.0278(62)&0.0191(49)&0.0262(91)&0.0240(38)\\
\hline
\multicolumn{6}{|c|}{Matrix Elements with (8,8) Operator} \\
\hline
&$\Delta t=20$&$\Delta t=24$&$\Delta t=28$&$\Delta t=32$&Error Weighted Ave.\\
\hline
$p=0$&0.0864(34)&0.0917(37)&0.0863(40)&0.0947(45)&0.0895(28)\\
$p=\pi/L$&0.0898(40)&0.0918(41)&0.0861(51)&0.0951(57)&0.0906(31)\\
$p=\sqrt{2}\pi/L$&0.0895(69)&0.0912(76)&0.087(11)&0.113(15)&0.0931(61)\\
$p=\sqrt{3}\pi/L$&0.092(19)&0.098(19)&0.120(32)&0.171(43)&0.111(14)\\
\hline
\multicolumn{6}{|c|}{Matrix Elements with (8,8) mixed Operator} \\
\hline
&$\Delta t=20$&$\Delta t=24$&$\Delta t=28$&$\Delta t=32$&Error Weighted Ave.\\
\hline
$p=0$&0.297(11)&0.315(12)&0.295(13)&0.324(15)&0.3068(91)\\
$p=\pi/L$&0.319(14)&0.327(14)&0.302(17)&0.333(19)&0.320(11)\\
$p=\sqrt{2}\pi/L$&0.336(25)&0.342(27)&0.319(37)&0.416(51)&0.347(21)\\
$p=\sqrt{3}\pi/L$&0.377(68)&0.385(74)&0.45(14)&0.66(15)&0.437(51)\\
\hline
\end{tabular}
\end{center}
\end{table}


\vspace{-0.15 in}
\section{Conclusion}
\vspace{-0.1 in}

Lattice matrix elements were computed for $\Delta I=3/2$, $K\to\pi\pi$ decays on quenched $24^3\times 64$ lattices, with a pion mass of $m_\pi=227.6(6)\text{ MeV}$, a kaon mass of $m_K=564(2)\text{ MeV}$, and pion momentum $p_\pi=\sqrt{n}\pi/L=170\sqrt{n}\text{ MeV}$ where $n$ is the number of twists (0, 1, 2, or 3).  It was necessary to reduce to the time separation between the kaon and the two pion sources to 20, 24, 28, and 32 units.  An error weighted average of the results from all four separations produced an overall result with a smaller error.  This error was of order 3-4\% for pion momentum 0 and $\pi/L$, 7\% for pion momentum $\sqrt{2}\pi/L$, and 15\% for pion momentum $\sqrt{3}\pi/L$.

Future plans include an expansion of this study (on the same lattices) to include a larger range of pion and kaon masses in order to perform an extrapolation to the physical pion and kaon mass, and to the physical (on-shell) pion momentum using the range of momenta obtained from the different twists.  The Lellouch-Luscher factor \cite{Lellouch_Luscher} will be used to convert the lattice matrix elements to physical decay amplitudes.  The calculation will also be performed on the RBC/UKQCD $32^3\times 64$, $L_s=32$ lattices with 2+1 flavors of DWF that are currently being generated, with a pion mass of $m_\pi\le 250\text{ MeV}$, in order to obtain results in the full dynamical theory which can be compared with the results of the quenched approximation.

\vspace{-0.15 in}
\section{Acknowledgements}
\vspace{-0.1 in}

Thanks to Tom Blum, Norman Christ, Chris Dawson, Chulwoo Jung, Changhoan Kim, Qi Liu, Robert Mawhinney, Chris Sachrajda, and all of our colleagues in the RBC and UKQCD collaborations for helpful discussions and the development and support of the QCDOC hardware and software infrastructure which was essential to this work.  In addition we acknowledge Columbia University, RIKEN, BNL and the U.S. DOE for providing the facilities on which this work was performed.  This work was supported in part by U.S. DOE grant number DE-FG02-92ER40699.  EG is supported by an STFC studentship and grant ST/G000557/1 and by EU contract MRTN-CT-2006-035482 (Flavianet).

\end{document}